\documentclass[prl,reprint,noshowpacs,superscriptaddress,floatfix,letterpaper]{revtex4-1}
\usepackage{amsmath,amssymb,amsbsy,amsfonts,amsthm}
\usepackage{bm,graphicx}
\usepackage{float}
\usepackage{color}
\usepackage{bm}
\usepackage{physics}
\usepackage[dvipsnames]{xcolor}
\usepackage[colorlinks=true,citecolor=Cerulean,linkcolor=RubineRed,urlcolor=Cerulean]{hyperref}
\usepackage{silence}
\usepackage{soul}
\usepackage[normalem]{ulem}

\newcommand\rsout{\bgroup\markoverwith{\textcolor{red}{\rule[0.5ex]{2pt}{0.8pt}}}\ULon}

\newcommand{\jrg}[1]{\textcolor{black}{#1}}

\newcommand{\llangle}{\langle\hspace{-.5mm}\langle\hspace{.2mm}}
\newcommand{\bllangle}{\bigg\langle\hspace{-1.6mm}\bigg\langle\hspace{.2mm}}
\newcommand{\rrangle}{\hspace{.2mm}\rangle\hspace{-.5mm}\rangle}
\newcommand{\brrangle}{\hspace{.2mm}\bigg\rangle\hspace{-1.6mm}\bigg\rangle}
\setstcolor{red}

\begin{document}

\title{Nonequilibrium uncertainty principle from information geometry}

\author{Schuyler~B.~Nicholson}
\affiliation{Department of Chemistry,\
 University of Massachusetts Boston,\
 Boston, MA 02125
}
\author{Adolfo~del~Campo}
\affiliation{Department of Physics,\
  University of Massachusetts Boston,\
  Boston, MA 02125
}
\affiliation{Center for Quantum and Nonequilibrium Systems,\
  University of Massachusetts Boston,\
  Boston, MA 02125
}
\author{Jason~R.~Green}
\email[]{jason.green@umb.edu}
\affiliation{Department of Chemistry,\
  University of Massachusetts Boston,\
  Boston, MA 02125
}
\affiliation{Department of Physics,\
  University of Massachusetts Boston,\
  Boston, MA 02125
}
\affiliation{Center for Quantum and Nonequilibrium Systems,\
  University of Massachusetts Boston,\
  Boston, MA 02125
}

\date{\today}

\begin{abstract}

With a statistical measure of distance, we derive a classical uncertainty
relation for processes traversing nonequilibrium states both transiently and
irreversibly. \jrg{The geometric uncertainty associated with dynamical histories
that we define} is an upper bound for the entropy production and flow rates,
but it does not necessarily correlate with the shortest distance to
equilibrium. For a model one-bit memory device, we find that expediting the
erasure protocol increases the maximum dissipated heat and \jrg{geometric}
uncertainty. A driven version of Onsager's three-state model shows that a set
of dissipative, high-uncertainty initial conditions, some of which are near
equilibrium, scar the state space.

\end{abstract}

\maketitle

Myriad phenomena generate structures and patterns that are unique outside of
thermodynamic equilibrium. Efforts to understand these processes stretch back
to the very beginnings of thermodynamics -- a pinnacle of physics that
encapsulates the quantitative understanding of energy transfer and
transformations~\cite{Callen85}. A powerful approach to studying thermodynamic
processes focuses on uncertainty
principles~\cite{mandelbrot1956outline,schlogl1988thermodynamic,uffink1999thermodynamic}.
Thermal uncertainty relations have strong resemblances to their quantum
counterparts and rest on the foundations of equilibrium statistical mechanics.
The recent introduction of nonequilibrium uncertainty
relations~\cite{barato2015thermodynamic,gingrich2016dissipation} has generated
a flurry of
activity~\cite{maes2017frenetic,horowitz2017proof,pietzonka2017finite,shiraishi2016universal,
shiraishi2017finite, proesmans2017discrete, dechant2017current}, but these
results are largely restricted to nonequilibrium steady-states. They leave open
the question of whether there are uncertainty relations for processes that are
transient and nonstationary. We address this question here.

There are growing links between thermodynamics and
information~\cite{Fluct_Orig,seifert2012stochastic,kawai2007dissipation,allahverdyan2009thermodynamic},
some of which place
bounds~\cite{hartich2014stochastic,horowitz2014thermodynamics,
yamamoto2016linear} on entropy changes~\cite{parrondo2015thermodynamics}. One
important example in this context is the erasure of physically-stored
information, which dissipates  heat and limits the computational power of
physical devices~\cite{Lloyd00}. There is still much to be done to disentangle
physical and logical irreversibility in order to clarify the processing of
information and thermodynamic function~\cite{boyd2016identifying}. Of
particular interest are extending predictions into practically important
regimes where erasure is fast and devices are small -- when dynamics and
statistical fluctuations rule. Progress in this direction requires a firm grasp
on the information in the distributions~\cite{Info_Theory} sampled by processes
driven transiently away from equilibrium.

For nonstationary processes, it is natural to treat the distributions evolving
under certain control parameters through ideas formalized in information
geometry~\cite{amari2007methods,Geo_Info,heseltine2016novel,nicholson2016structures,oizumi2016unified}.
\jrg{There, the focus is on the structure of the manifold of probability
distributions, along with the distance and velocity paths traversed by the
dynamics of the system.} Though often presented in a general
setting~\cite{amari2007methods}, information geometry has connections to
thermodynamics~\cite{LthOriginal,MLthermo,LthSorig,sivak2012thermodynamic,lahiri2016universal}.
For nonstationary irreversible processes, results are scarce, however, and our
understanding between thermodynamics and information geometry remains
incomplete. A significant challenge to the development of a
statistical-mechanical theory for nonstationary processes is that there are few
restrictions on the possible nonequilibrium distributions over paths or states.
In this Letter, we establish a fundamental connection  between the acceleration
of the Shannon entropy and the Fisher information that enables us to bring the
mathematical machinery of information geometry to bear on the problem.

\textit{Notation and setting.--} At the ensemble level, a path is the set of
probability distributions a system samples as it evolves over a finite time
interval. We define the set of probability distributions $\mathbb{P}(\Omega) =
\{p:\Omega \to \mathbb{R}\, |\, p_x(t) > 0 , \forall x \in  \Omega,  \sum_x
p_x(t) = 1\}$. A subset of these distributions belong to the manifold $\Theta =
\left\{p(x|\theta(t)): \theta(t) =
\{\theta^1(t),\theta^2(t),\ldots,\theta^N(t)\}\right\}$, where $\theta(t)$
represents the time-dependent control parameters~\cite{amari2007methods}
determining the path across the manifold. Empirically, one could sample
trajectories through the system state space and construct a distribution at
each moment in time from the ensemble of realizations (each distribution being
a point on $\Theta$ in the large sample limit). Together, these distributions
are the ``path'' in probability space between the initial distribution,
$p(t_0)$, and the final distribution, $p(t_f)$, over the time interval $\tau =
t_f - t_0$. Here, we assume no particular form for the distributions and
instead consider the system dynamics governed by the master equation
\begin{equation}
  \dot{p}_x(t) = \sum_y W_{xy}(\theta(t))p_y(t),
  \label{eq:ME}
\end{equation}
where $\dot{p}_x(t) = dp_x(t)/dt$ and $W_{xy}(t)$ is the transition rate from
state $y \rightarrow x$. The occupation probability $p_x(t) = p(x|\theta(t))$
for state $x$ is conditional on the control parameters $\theta(t)$. The rate
matrix $\mathbb{W}(t)$ also depends on $\theta(t)$ and follows the usual
conventions: for $W_{xy}(t) \in \mathbb{W}(t)$, $W_{xy}(t)> 0$ when $x \neq y$
and $W_{yy}(t) = -\sum_{x\neq y} W_{xy}(t)$ so that $\sum_x W_{xy}(t) = 0$. 

A system satisfies detailed balance if the currents or thermodynamic fluxes,
$C_{xy}(t) = W_{xy}(t)p_y(t) - W_{yx}(t)p_x(t)$, are zero for all $x,y$.
Otherwise, the existence of current implies the system is undergoing an
irreversible process~\cite{schnakenberg1976network}. The current is related to
the dynamics through the master equations, $\dot p_x(t) = \sum_y C_{xy}(t)$.
But, it does not satisfy the requirements of a metric and, so, cannot be used
to quantify the distance from equilibrium. \jrg{However, it is well known that
the Fisher information is a metric~\cite{Rao1945}, providing a notion of
distinguishability between neighboring distributions related by the
time-evolution of the dynamics. Here, we arrive at the Fisher information and
the ``geometric uncertainty'' accumulated along a path across $\Theta$ through
the matrix,}
\begin{align}
  E_{xy}(t) &= W_{xy}(t) - \frac{C_{xy}(t)}{2p_y(t)}. 
  \label{eq:EDef1}
\end{align}
The results that follow are built on the foundation set by the properties of
this matrix (see Supplemental Material (SM)). Even when the current is nonzero,
this matrix satisfies a detailed balance condition, $E_{xy}(t)p_y(t) =
E_{yx}(t)p_x(t)$. It is similar to a symmetric matrix and, thus, has a complete
set of eigenvectors and real eigenvalues~\cite{horn2012matrix}.  Matrices with
a similar form and function are known for discrete-time, discrete-state Markov
chains~\cite{NOR,nicholson2016geometric} but not for continuous-time Markovian
dynamics. As we will show, $E$ allows us to connect the Fisher information
(from information geometry) to the entropic acceleration (from thermodynamics).

\textit{Fisher information and thermodynamics.--} The Fisher
matrix~\cite{amari2007methods},
\begin{equation}
  g_{ij} = \sum_{x}p_x(t)\frac{\partial\ln p_x(t)}{\partial\theta_i}\frac{\partial\ln p_x(t)}{\partial\theta_j},
\end{equation}
is a metric tensor that gives a statistical measure of distance over a manifold
of probability distributions, $ds^2 = \sum_{i,j} g_{ij}d\theta_id\theta_j$.
The Fisher information, $I_F(t)$, reflects a change in a probability distribution
with respect to a set of control parameters~\cite{FisherInfo}. When
parametrized by time it is
\begin{align}
  I_F(t) = \sum_{i,j}\frac{d\theta_i}{dt}g_{ij}\frac{d\theta_j}{dt}
  = \sum_{x}p_x(t)\left[\frac{d\ln p_x(t)}{dt}\right]^2.
  \label{eq:IFdef}
\end{align}
Thus far, the Fisher information is purely a mathematical construction.
However, we can relate it to the acceleration of the entropy through the
entropy production/flow and thereby add to the known connections between
information and thermodynamics.

Shannon~\cite{Shannon} showed that $I_y(t) = -\ln p_y(t)$ is the information
associated with state $y$. The difference $I_{xy}(t) = -\ln p_y(t)/p_x(t)$ is then
the local difference in information or ``surprise'' between state $y$ and state
$x$. With this context, consider the Shannon entropy entropy rate,
\begin{align}
  \dot{S}(t) &\equiv -\sum_x \dot{p}_x(t)\ln p_x(t) = -\llangle I(t) \rrangle, 
\end{align}
which we express as an average, $\llangle \cdot \rrangle \equiv
\sum_{x,y}W_{xy}(t)p_y(t)[\cdot]$, that is equivalent to an average over the
current (up to a factor of $1/2$)~\cite{esposito2010three}. The connection to
nonequilibrium thermodynamics comes from decomposing the entropy rate at any
instant in time, $\dot{S} = \dot{S}^i  + \dot{S}^e$, into the entropy production rate from sources in the system, $\dot{S}^i$, and the rate of entropy exchange with the environment, $\dot{S}^{e} =
-\sum_{x,y}W_{xy}(t)p_y(t)\ln W_{xy}(t)/W_{yx}(t)$~\cite{seifert2005entropy}.
The entropy production $\dot{S}^i  = \llangle F\rrangle$ is an average of the
generalized forces, $F_{xy} = \ln W_{xy}(t)p_y(t)-\ln
W_{yx}(t)p_x(t)$~\cite{esposito2010three}, which multiplied by Boltzmann's
constant, $k_B$, are the thermodynamic
affinities~\cite{schnakenberg1976network}. Here, we set $k_B = 1$.

The second derivative of the Shannon entropy is the ``entropic acceleration'',
\begin{equation}
  \ddot{S}(t) = -\frac{d}{dt}\llangle I(t)\rrangle = -\sum_x \ddot{p}_x(t)\ln p_x(t) - \llangle\dot{I}(t)\rrangle.
  \label{eq:ddotS}
\end{equation}
Our first main result is that this acceleration relates to the Fisher
information, which for nonstationary irreversible Markovian processes is an
average over the rate of information change in the system,
\begin{align}
  I_F(t) 
  &= -\sum_{x,y}W_{xy}(t)p_y(t) \frac{d}{dt}\ln\left[\frac{E_{yx}(t)}{E_{xy}(t)}\right]
  = \llangle \dot{I}(t)\rrangle.
  \label{eq:IF1}
\end{align} 
Combining Eq.~(\ref{eq:ddotS}) and (\ref{eq:IF1}), shows that the Fisher
information and the entropic acceleration are related:
\begin{align}
  \ddot{S}(t) &= -\sum_x \ddot{p}_x(t)\ln p_x(t) - I_F(t)
  =\ddot{S}^i + \ddot{S}^e.
  \label{eq:SIF}
\end{align}
This result can be cast in matrix form with $E(t)$ (SM), which can also be
expressed in terms of the thermodynamic forces. The entropic acceleration
measures the rate at which the bulk information changes in time, the Fisher
information is the local rate of information change on average, and the
remainder is their sum, $\mathcal{C} = -\sum_x \ddot{p}_x(t)\ln p_x(t) = d\llangle I
\rrangle/dt + \llangle dI/dt \rrangle$.

\begin{figure*}[!ht]
\centering
\includegraphics[width=.8\textwidth]{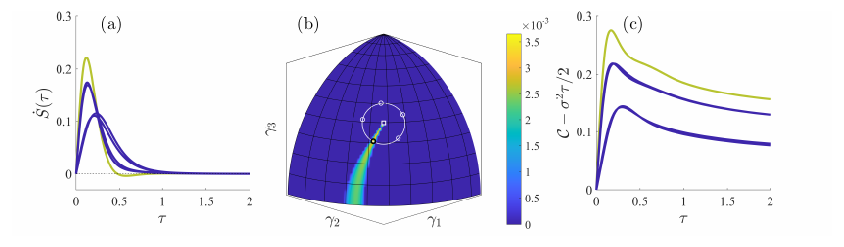}

\caption{\label{fig:SurfFig} Scars on the state space (a)-(c) emerge over time
from initial conditions that ultimately dissipate and have higher path
uncertainties. The entropy rate $\Delta \dot{S} = \dot{S}(t_f)$ as a function
of time $\tau$ for five initial conditions equidistant from equilibrium (colors
correspond to (b)). The initial conditions are marked by circles in (b), which
shows the positive octant of a sphere colored by $\dot{S}(t_f)$ when the path
reaches the stationary distribution (white square). At $t=0$, $\dot{S}(t_0) =
0$. (c) The upper bound for $\dot{S}(t)$, $\mathcal{C}-\sigma^2\tau/2$, for the
five select initial conditions as a function of time.}

\vspace{-0.1in}

\end{figure*}

\textit{Uncertainty and deviations from the geodesic.--} Now, by introducing a
measure of uncertainty over a path across $\Theta$, \jrg{these results enable
us to show that for any initial and final distribution, the entropy rate is
bounded from above by the contributions from the local and bulk information
rates and the geometric uncertainty about the path. Rao showed the Fisher
information matrix satisfies the requirements of a metric~\cite{Rao1945}} and,
so, the Fisher information relates to the line element between two
distributions infinitesimally displaced from one another, $ds^2 =
I_F(t)\,dt^2$. The length, $\mathcal{L}$, of a path on the manifold $\Theta$
can then be measured with the statistical distance~\cite{StatDistQuantum},
$\mathcal{L} = \int_{t_0}^{t_f} dt \sqrt{I_F(t)}$. The Cauchy-Schwarz
inequality yields the statistical divergence,
\begin{equation}
  \mathcal{J} \equiv \tau\int_{t_0}^{t_f}dt\,I_F(t) \geq \mathcal{L}^2.
  \label{eq:JDif}
\end{equation}
Previous work has shown that $\mathcal{J} - \mathcal{L}^2 \geq 0$ is a temporal
variance~\cite{heseltine2016novel,nicholson2016structures}, and that, in one
representation, can measure cumulative fluctuations in the rate coefficients
for irreversible decay processes~\cite{flynn2014measuring,*nichols2015order}.

\jrg{In the current context, the difference between the two terms of the
Cauchy-Schwarz inequality equals the variance or ``geometric uncertainty'' of the path
connecting $p(t_0)$ and $p(t_f)$ that measures the cumulative deviations from
the geodesic.} To illustrate this interpretation, we define the time average
for a function, $A(t)$, as $\mathbb{E}[A(t)] = \tau^{-1}\int_{t_0}^{t_f}
dt\,A(t)$. The difference between the time average of $I_F$ and the squared
time-average of $\sqrt{I_F}$ over the path is the time-averaged variance
\begin{align}
  \sigma^2
  = \frac{\mathcal{J} - \mathcal{L}^2}{\tau^2} = \mathbb{E}\left[I_F \right] - \mathbb{E}\left[\sqrt{I_F}\right]^2 \geq 0. 
  \label{eq:IrrM}
\end{align}
\jrg{This geometric uncertainty is the cumulative deviation from the geodesic
connecting the initial and final distributions. It} depends on the path and the
initial and final distributions. We expect it to be nonzero for most
irreversible processes. One notable exception are paths following the geodesic
connecting two distributions. These paths correspond to the condition
$\mathcal{J} = \mathcal{L}^2$~\cite{StatDistQuantum} and a variance of zero.
These ``certain'' paths are irreversible, nonstationary paths with zero
\jrg{geometric} uncertainty.

It has previously been shown that measuring cumulative deviations from the
geodesic amounts to measuring the cumulative fluctuations in nonequilibrium
observables~\cite{flynn2014measuring,nichols2015order,sivak2012thermodynamic}.
Past work has also used statistical distances (though with other metrics) to
measure the dissipation associated with quasistatic
transformations~\cite{Dissipated_Availability}. These results, however, do not
connect thermodynamic quantities such as the entropic acceleration to the
Fisher information for general nonstationary irreversible processes as we do
here.

\jrg{Our second main result is a bound on the entropy rate by the geometric
uncertainty.} It follows from recognizing that the variance satisfies the
inequality:
\begin{align}
  \sigma^2
  &\leq \frac{1}{\tau}\int_{t_0}^{t_f} dt \left(I_F + \mathbb{E}\left[\sqrt{I_F}\right]^2\right) \leq \frac{2\mathcal{J}}{\tau^2}.
  \label{eq:sigInq}
\end{align}
The last step uses $\mathcal{J} - \mathcal{L}^2 \geq 0$ and the nonnegativity
of the variance. Defining $\mathcal{I} \equiv \tau^{-1}\int_{t_0}^{t_f} dt\,
\llangle \dot{I} \rrangle$, this relation becomes
\begin{align}
  \mathcal{I}\,\sigma^{-2} \geq \frac{1}{2}.
  \label{eq:IFsOne}
\end{align}
The intuition behind this uncertainty relation is that \jrg{different paths
across the manifold of probability distributions $\Theta$ can lower the
time-averaged rate of information change $\mathcal{I}$, but only at the expense
of a corresponding decrease in uncertainty (a smaller excursion from the
geodesic).} Simply put, the uncertainty places a bound on the cumulative rate
of information change. It is worth noting that this information-uncertainty
ratio is valid for nonstationary, irreversible paths over any finite time
interval between arbitrary probability distributions. \jrg{To test this
inequality, one only needs the basic ingredients of a Markov state model,
models that have proven useful for discovering collective variables and
analyzing rare events in diverse areas, including protein
(un)folding~\cite{LiK13,HusicP18}.}

Another way to write the uncertainty relation is in terms of the entropic
acceleration, Eq.~(\ref{eq:SIF}). Upon integrating, it becomes a bound on the
entropy rate
\begin{equation}
  \Delta\dot{S} \leq \mathcal{C} - \frac{\sigma^2\tau}{2},
  \label{eq:SdotSig}
\end{equation}
where, again, $\mathcal{C} = -\int_{t_0}^{t_f}dt\sum_x\ddot{p}_x(t)\ln p_x(t)$
and $\Delta\dot{S} = \dot{S}(t_f) - \dot{S}(t_0)$. \jrg{The quantities
$\mathcal{C}$ and $\sigma^2\tau/2$ can both be zero only in a stationary state
-- that is, this uncertainty relation applies specifically to the nonstationary
regime -- and are measurable from occupation/transition probabilities}. A
direct connection to previous stationary uncertainty relations appears to be a
subtle question. \jrg{However, the present result has a clear physical meaning:
the entropy rate (from thermodynamics) is bounded by contributions from the
local and bulk information dynamics and cumulative deviations from the geodesic
(from information geometry).} When no heat is exchanged, these information
dynamics bound the entropy production. And, when there are no internal sources
of entropy production, they bound the entropy (heat) flow.

\jrg{The results so far avoid any assumptions about the probability
distributions, rate of driving, or ``closeness'' to equilibrium. For additional
insight into the bounds placed on energy exchange, consider a system in contact
with a heat bath at fixed temperature $T$, in which the energy of each state of
the system is driven slowly.} At each moment in time, the probability of state
$x$ is $p_x(t) = e^{-\beta\epsilon_x(t)}/Z(t)$, where $\epsilon_x(t)$ is the
energy of state $x$ and $Z(t)$ is the partition function. Eq.~(\ref{eq:IFsOne})
becomes $\tau^{-1}\int_{t_0}^{t_f}\textrm{Var}\left[\dot{\epsilon}(t)\right]dt
\geq \sigma^2/2\beta^2$. \jrg{The uncertainty measures the cumulative
deviations from constant energy rate fluctuations. The geodesic corresponds to
a path where the energy rate fluctuations are time independent.
During a nonstationary process operating near this bound, lowering the
time-average fluctuations in energy flux will mean smaller excursions from the
geodesic where these fluctuations and $I_F(t)$ are constant:
$I_F(t)=\beta^2\textrm{Var}\left[\dot{\epsilon}(t)\right]\geq 0$.}

\textit{Uncertainty scarring in a single-cycle chemical reaction.--} To
illustrate these results, we adapt the kinetic scheme used by
Onsager to demonstrate the reciprocal relations of irreversible
thermodynamics~\cite{onsager1931}. The model consists of three states and a
kinetics driven by the time-dependent rate coefficients, $k^+ = 4\atan(\omega_1
t)$ and $k^- = 4\atan(\omega_2 t)$, with $\omega_1 = 4$ and $\omega_2=6$. The
inverse tangent function ensures that for large $t$, every path reaches the
same stationary distribution, $p_x^{\infty}=(1/3,1/3,1/3)$. Our criterion for a
path to reach the stationary distribution is that each initial condition must
evolve to be within $\Vert p_x(t) - p_x^{\infty} \Vert_2 \leq 5\times 10^{-3}$
of the stationary distribution. Under the transformation $\gamma_x(t) \equiv
\sqrt{p_x(t)}$, the system travels across the positive octant of a sphere
(Fig.~(\ref{fig:SurfFig})).

This system and driving protocol localize the effects of the initial condition
on the \jrg{geometric uncertainty} about the nonequilibrium path (SM).  What we find is
that the distance from the stationary state says little about the uncertainty
or the entropy rate (dissipation rate).  Figure~(\ref{fig:SurfFig}b)
shows $\Delta \dot{S}=\dot{S}(\tau)$ for all physically-relevant initial
conditions (color indicates final $\dot{S}$). Five initial conditions are
marked (open circles), each point equidistant from the equilibrium state,
$p_x^\infty$ (white square). While these initial conditions are all equally
``far'' from equilibrium, their entropy rates exhibit different behavior over
time (Figure~(\ref{fig:SurfFig}a), color corresponds to those in (b)).
Moreover, paths originating on the ``scar'' (Fig.~(\ref{fig:SurfFig}b)), have a
larger maximum value of $\dot{S}(\tau)$ than those launched from off it.  These
initial conditions are also unique in that they all dissipate for a period of
time. Initial conditions off the scar do not dissipate. Regardless of the
dissipative nature of these paths, the uncertainty relation holds. The time
dependence of the upper bound is shown in Fig.~(\ref{fig:SurfFig}c) for the
same five initial conditions. Again, those initial conditions originating in
the scar have a larger uncertainty and upper bound the entropy rate.  Overall,
these results are evidence that the distance from the stationary state can be a
poor predictor of transient nonequilibrium behavior.

\begin{figure}
\centering
\includegraphics[width = 1\columnwidth]{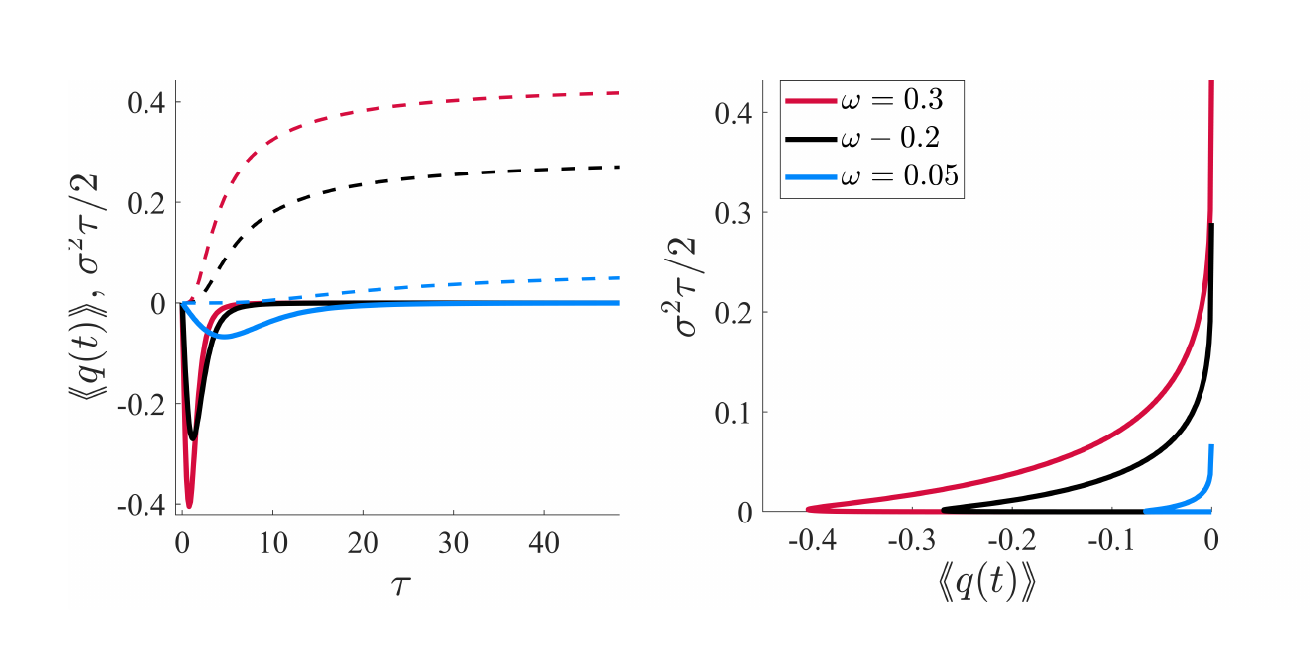}

\caption{\label{fig:Land} Accelerating the erasure of information (destroying
state $x = 0$), by increasing $\omega$, generates more (a) uncertainty
($\sigma^2\tau/2$, dashed lines) about the erasure path and maximum heat
dissipated ($\llangle q(t)\rrangle$, solid lines).  (b) The heat dissipated,
$\llangle q(t)\rrangle$, follows the geodesic ($\sigma^2\approx 0$) up until
the maximum amount of heat is being released.  Beyond that point, the system
undergoes entropic deceleration. Time series are from $p(t_0)= [0.5,0.5]$ to
$p(t_f)= [1,0]$ for $\omega = [0.05,0.2,7=0.3]$ shown in blue, black, and red
lines, respectively.}

\vspace{-0.2in}

\end{figure}

\textit{Landauer's principle and information erasure.--} Perhaps nowhere is the
connection between information and thermodynamics more apparent than in
Landauer's principle~\cite{Landauer61}. According to this principle, erasing
one bit of information requires the dissipation of at least $k_BT\ln 2$ thermal
energy as heat. Since the bound on $\Delta \dot{S}(t)$ in
Eq.~(\ref{eq:SdotSig}), holds for any Markovian evolution between any two
distributions, we can explore the connection between the entropy (heat)
dissipation and the uncertainty about the erasure path. \jrg{To examine
correlation between erasure paths and the heat release, we consider the erasure
of one-bit of information in a model memory device.}

The model 1-bit memory device initially consists of two states $x = \{1,0\}$
that are equally probable, $p(t_0) = [0.5,0.5]$. To measure the heat release,
we choose $W(t)$ such that every distribution is of the form $p_x(t) =
Z(t)^{-1}e^{-\beta \epsilon_x(t)}$, where $Z(t)$ is the partition function,
$\beta$ is the inverse temperature set to one, and $\epsilon_x(t)$ is the
energy at state $x$ (SM).  The energy at $x=1$, is held fixed and
$\epsilon_2(t) = c_1 + c_2/\pi \atan(\omega t)$ depends on $\omega$. Here, $c_1
= 0.2$ and $c_2 = 20$. The parameter $\omega$, controls the rate at $k_B T\ln
2$ of energy is dissipated. This restriction on $p_x(t)$ means that the entropy
rate, $\dot{S} = -\sum_{x,y}W_{xy}p_y(\epsilon_x(t) - \epsilon_y(t)) = \llangle
q(t)\rrangle$, is the average heat exchange between the system and surroundings
at an instant in time. From Eq.~(\ref{eq:SdotSig}), we know that $\llangle
q(t)\rrangle \leq \mathcal{C} - \sigma^2\tau/2$. For this system, both
$\mathcal{C}$ and $\sigma^2\tau/2$ are positive and $\llangle q(t)\rrangle \leq
0$, as expected. The higher the rate at which heat is dissipated, the larger
the erasure path uncertainty (Fig.~(\ref{fig:Land}a)). The uncertainty does not
increase at a constant rate during the erasure protocol. As energy is initially
dissipated, up to the maximum value, the system approximately follows the
geodesic across $\Theta$ and $\sigma^2 \approx 0$. Figure~(\ref{fig:Land}b)
shows $\sigma^2\tau/2$ is near zero until the system reaches a state of maximal
dissipation, after which the path moves off the geodesic. The faster
physically-stored information is erased, the faster energy is dissipated and
the greater the resulting uncertainty about the path to equilibrium.

\textit{Conclusions.--} For processes arbitrarily far from equilibrium, we have
established a bound on the entropy production and flow rates via the
uncertainty in the  path connecting any two arbitrary distributions. This
uncertainty relation holds when the system evolves under a time-inhomogeneous
Markovian dynamics, making it applicable to a broad class of nonequilibrium
processes. It is clear that even for the classical single-cycle system, the
proximity to the stationary state is a poor indicator of uncertainty: initial
conditions that are statistically equidistant from the stationary state can
have dramatically different geometric uncertainties, uncertainties that we
showed are linked to the entropy rate.  When erasing information in a model
one-bit memory device, we find that increasing the speed of erasure comes at
the expense of increasing the rate of energy dissipation and the geometric
uncertainty about the path to equilibrium. We expect these results to be
usefully applied to other kinetic phenomena, such as (bio)chemical
reactions~\cite{barato2014efficiency,mcgrath2017biochemical}, and further
expand the understanding of processes away from equilibrium, both near and far.

\textit{Acknowledgments.--} This material is based upon work supported by the
U.S. Army Research Laboratory and the U.S. Army Research Office under grant
number W911NF-14-1-0359 and the John Templeton Foundation. S.~B.~N.\
acknowledges financial support from the Office of Global Programs, University
of Massachusetts Boston. We thank Sosuke Ito for pointing out the relation
between the generalized forces and the matrix $E$ and Tamiki Komatsuzaki for
comments on the manuscript.

\vspace{-0.1in}


%


\section*{Supplemental Information}

\section{Properties of the matrix, $E$}
In this section, we will derive properties of the matrix,
\begin{equation}
  E_{xy}(t) = W_{xy}(t) - \frac{C_{xy}(t)}{2p_y(t)},
  \label{eq:EDef}
\end{equation}
that underlie the results in the main text. We will use bra and ket notation
$\bra{\cdot}\ket{\cdot}$, so that given the $N\times N$ matrix, $E$, we have
$[E\ket{p}]_x = \sum_y E_{xy}p_y$ and $[\bra{p}E]_y = \sum_x p_x E_{xy}$.\\

\noindent \textbf{I\@. Master equation.} The master equation can be recast in
terms of the matrix $E(t)$. Multiplying $E(t)$ by the probability $p_y(t)$ and
summing gives,
\begin{align}
  \sum_y E_{xy}(t)p_y(t)
  &= \sum_y\left( W_{xy}(t) - \frac{C_{xy}(t)}{2p_y(t)} \right)p_y(t), \nonumber \\
  \sum_y E_{xy}(t)p_y(t) &= \dot{p}_x(t) - \frac{\dot{p}_x(t)}{2}, \nonumber \\
  2\sum_y E_{xy}(t)p_y(t) &= \dot{p}_x(t)
  = \sum_y W_{xy}(t)p_y(t).
  \label{eqn:eme}
\end{align} 
The final line shows that $E(t)$ is an alternative representation of the
dynamics governed by the master equation.

We should note that $E(t)$ is not a trivial transformation of $W(t)$. For
example, $2E(t) \neq W(t)$. Instead, multiplying $E(t)$ by two, shows that if
$C_{xy}(t) = 0$ $\forall\,x,y$ then $E(t) = W(t)$. In general, however,
$2E_{xy}(t) = W_{xy}(t) + \hat{W}_{xy}(t)$, where $\hat{W}_{xy}(t)$ is the
time-reversed rates defined~\cite{Markov_Chains} by
\begin{align}
  \hat W_{xy}(t)p_y(t) &= W_{yx}(t)p_x(t), \nonumber \\
  \hat W_{xy}(t) &= \frac{1}{p_y(t)}W_{yx}(t)p_x(t).
\end{align}
The matrix $\hat W(t)$ defines the microscopically reverse dynamics of
$W_{xy}(t)$, which do not satisfy detailed balance in general. In fact, only
when $W(t)$ satisfies detailed balance, does $W_{xy}(t)p_y(t)  = \hat W_{xy}(t)
p_y(t)$.

\bigskip

\noindent\textbf{II\@. Detailed balance condition.} The rate matrix has the
properties that $W_{xy}(t) \in \mathbb{W}(t)$, $W_{xy}(t) > 0$ for $x \neq y$
and $W_{yy}(t) = -\sum_{x\neq y} W_{xy}(t)$ so that $\sum_x W_{xy}(t) = 0$.
Using the last property, the master equation becomes
\begin{align}
  \dot{p}_x(t) &= \sum_{y\neq x} \left[W_{xy}(t)p_y(t) - W_{yx}(t)p_x(t)\right]\nonumber\\
  &= \sum_y C_{xy}(t).
  \label{eqn:me}
\end{align}
The master equation exhibits detailed balance if each of the currents vanish;
that is, when the currents or thermodynamic fluxes, $C_{xy}(t) =
W_{xy}(t)p_y(t) - W_{yx}(t)p_x(t)$, are zero for all $x,y$.  Otherwise, the
existence of current implies the system is undergoing an irreversible
process~\cite{schnakenberg1976network}.

Even for processes that are driven or transiently away from equilibrium,
$C_{xy}(t)\neq 0$, the matrix $E(t)$ satisfies a similar detailed balance
condition. Since the master equation can be recast in terms of $E(t)$, it also
has an analogous form with source and sink terms. Applying the definition of
$E(t)$ to the master equation gives
\begin{align}\nonumber
  \dot{p}_x(t) &= \sum_{y\neq x} \left[E_{xy}(t)p_y(t) - E_{yx}(t)p_x(t) + C_{xy}(t)\right]\\
  &= \sum_y \left[C_{xy}^E(t) + C_{xy}(t)\right].
\end{align}
In the final line, we define the current (for $E(t)$) between states $x$ and $y$,
$C_{xy}^E(t)$. Comparing this result to Eq.~(\ref{eqn:me}) suggests
$C_{xy}^E(t)=0$, akin to detailed balance, but valid when $C_{xy}(t)\neq 0$. To
prove this detailed balance condition, we can expand the current for $E(t)$: 
\begin{align}
  C^E_{xy}(t) &\equiv E_{xy}(t)p_y(t) - E_{yx}(t)p_x(t) \nonumber\\\nonumber
  & = W_{xy}(t)p_y(t) - \frac{C_{xy}(t)}{2} - W_{yx}(t)p_x(t) + \frac{C_{yx}(t)}{2} \\\nonumber
  &= \frac{C_{xy}(t)}{2} + \frac{C_{yx}(t)}{2} \\
  &= 0.
\end{align}
The last equality follows from the anti-symmetry of the current, $C_{xy}(t) =
-C_{yx}(t)$. When detailed balance is satisfied for $C_{xy}(t)$ it is also
satisfied for $C_{xy}^E(t)$: the condition for both is $W_{xy}(t)p_y(t) =
W_{yx}(t)p_x(t)$.\\

\noindent\textbf{III\@. Symmetrization.} As is done at equilibrium with $W(t)$,
we can show that $E(t)$ is similar to a symmetric matrix, $S(t)$: $E(t)
\backsim S(t)$. First, we define
\begin{equation}
  S_{xy}(t) = \frac{1}{\sqrt{p_x(t)}}E_{xy}(t)\sqrt{p_y(t)}.
  \label{eq:SDef}
\end{equation}
To show $S(t)$ is symmetric, we use $S(t)$ and the detailed balance of $E(t)$:
\begin{align}\nonumber
  E_{xy}(t)p_y(t) &= E_{yx}(t)p_x(t), \\\nonumber
  \sqrt{p_x(t)}S_{xy}(t)\sqrt{p_y(t)} &= \sqrt{p_y(t)}S_{yx}(t)\sqrt{p_x(t)}, \\
  S_{yx}(t) &= S_{xy}(t).
\end{align}
Since $S(t)$ is a real, symmetric (Hermitian) matrix, it has a complete set of
eigenvectors and real eigenvalues~\cite{horn2012matrix}. Since $E(t)$ is
similar to $S(t)$, it also has a complete set of eigenvectors and real
eigenvalues. We note that the eigenvectors and eigenvalues of both matrices are time
dependent.

The $E$-representation of the master equation, Eq.~(\ref{eqn:eme}), does not
imply that $W(t)$ is similar to $S(t)$. From the definitions of $E(t)$ and
$S(t)$, Eq.~(\ref{eq:EDef}) and Eq.~(\ref{eq:SDef}), we know
\begin{equation*}
  E_{xy}(t) = \sqrt{p_x(t)}S_{xy}(t)\frac{1}{\sqrt{p_y(t)}}.
\end{equation*}
Re-writing in terms of $S(t)$ gives:
\begin{equation*}
  S_{xy}(t) = \frac{1}{\sqrt{p_x(t)}}W_{xy}(t)\sqrt{p_y(t)} - \frac{C_{xy}(t)}{2\sqrt{p_x(t)p_y(t)}}.
\end{equation*}
By inspection, $W(t)$ is only similar to $S(t)$ when $C_{xy}(t) = 0$ $\forall$
$x,y$ and $E(t) = W(t)$. It can be further shown that just because $E(t)
\backsim S(t)$ through $p_x(t)$, this does not imply $\exists$ $S'(t) \backsim
W(t)$, where $S'_{xy}(t) = S'_{yx}(t)$. Defining $S'(t)$  as
\begin{equation*}
  S'_{xy}(t) = \frac{1}{\sqrt{p_x(t)}}W_{xy}(t)\sqrt{p_y(t)}.
\end{equation*}
and using the expression for the current,
\begin{align}
  C_{xy}(t) &= W_{xy}(t)p_y(t) - W_{yx}(t)p_x(t) \nonumber \\
  &= \sqrt{p_x(t)}S'_{xy}(t)\sqrt{p_x(t)} - \sqrt{p_y(t)}S'_{yx}(t)\sqrt{p_x(t)} \nonumber,
\end{align}
yields
\begin{equation*}
  S'_{xy}(t) = \frac{C_{xy}(t)}{\sqrt{p_x(t)p_y(t)}} + S_{yx}'(t).
\end{equation*}
Thus, $S'(t)$ cannot be symmetric unless the detailed balance is satisfied,
$C(t) = 0$, or $S'$ is the zero matrix. It is the nonstationary,
irreversibility of the system that prevents $W(t)$ from satisfying a similarity
transform -- irreversibility is built into $E(t)$.\\

\noindent \textbf{IV\@. Surprisal rate.} Shannon~\cite{Shannon} identified the
information gained or surprise in observing the state $y$ as $-\ln
p_y(t)$~\cite{Info_Theory}. Using $\sum_x E_{xy}(t) = [\bra{1}E(t)]_y$, where
$\bra{1}$ is a row vector of ones, the surprisal rate is related to $E$ by
\begin{align}
  [\bra{1}E(t)]_y &= \sum_x \left[W_{xy}(t) - \frac{C_{xy}(t)}{2p_y(t)}\right] \nonumber \\
  & = \frac{\dot p_y(t)}{2p_y(t)} = \frac{1}{2}\frac{d \ln p_y(t)}{dt}.
  \label{eq:EI}
\end{align}
This relationship also implies, through conservation of probability, that
$\bra{2} E(t)\ket{p(t)} = \sum_y \dot p_y(t) = 0$.\\

\noindent \textbf{V\@. Fisher information.} Underlying the principal results of
the main text is that $E(t)$ is related to the Fisher information, $I_F(t)$,
through $I_F(t) = -\bra{2}\dot{E}(t)\ket{p(t)} = \bra{2}E(t)\ket{\dot{p}(t)}$.
Differentiating $\bra{2} E(t)\ket{p(t)}$ with respect to time gives
\begin{equation}
  \frac{d}{dt}\bra{2} E(t)\ket{p(t)} = \bra{2}E(t)\ket{\dot{p}(t)} +
\bra{2}\dot{E}(t)\ket{p(t)}.
  \label{eq:EIdot}
\end{equation}
The first term on the right-hand side is the Fisher information, 
\begin{align}
  \bra{2}E(t)\ket{\dot{p}(t)} &= 2\sum_{x,y}\left( W_{xy}(t)\dot{p}_y(t) - \frac{C_{xy}(t)\dot p_y(t)}{2p_y(t)}\right)\nonumber \\
  &=  \sum_{x,y}\frac{1}{p_y(t)}C_{yx}(t)\dot p_y(t) \nonumber \\
  &= \sum_y \frac{\dot p_y(t)^2}{p_y(t)} \nonumber \\
  &= I_F(t).
  \label{eq:EIF1}
\end{align}
The second term is the negative of the Fisher information:
\begin{align}
  \bra{2}\dot{E}(t)\ket{p(t)} &= 2\sum_{xy} \left(\dot W_{xy}(t)p_y(t) \right. \nonumber\\
  & \quad\quad\left. \vphantom{} + \frac{C_{xy}(t)\dot p_y(t)}{2p_y(t)} - \frac{\dot{C}_{xy}(t)}{2}\right) \nonumber \\
  &= \sum_{x,y} \frac{C_{xy}(t)\dot p_y(t)}{p_y(t)}\nonumber \\
  &= -\sum_{x,y} \frac{C_{yx}(t)\dot p_y(t)}{p_y(t)}\nonumber \\
  &= -I_F(t).
  \label{eq:EIF2}
\end{align}
From the first to the second line, we use conservation of probability,
$d/dt\sum_x W_{xy}(t) = 0$, and $d/dt \sum_{xy} C_{xy}(t) = 0$. Plugging
Eq.~(\ref{eq:EIF1}) and Eq.~(\ref{eq:EIF2}) into Eq.~(\ref{eq:EIdot}), we see
that the conservation of probability leads to $\frac{d}{dt}\bra{2}
E(t)\ket{p(t)} = I_F(t) - I_F(t) = 0$.\\

\noindent \textbf{VI\@. Fisher information as entropic acceleration.} With the
properties discussed so far, we can arrive at the first main result: for systems with
dynamics that are governed by continuous-time master equations, the Fisher
information is part of the entropic acceleration. To show this, we recognize that
the matrix $E(t)$ in Property~IV can be expressed in terms of the generalized
thermodynamic forces, 
\begin{align}
  \left[\bra{2}E(t)\,\right]_y &= \frac{1}{p_y(t)}\sum_x C_{yx}(t)\nonumber \\
  &= \frac{1}{p_y(t)}\sum_x W_{yx}(t)p_x(t) - W_{xy}(t)p_y(t) \nonumber \\
  &= \frac{1}{p_y(t)}\sum_x W_{xy}(t)p_y(t)\left(\frac{W_{yx}(t)p_x(t)}{W_{xy}(t)p_y(t)} - 1\right)\nonumber \\
  &= \sum_x W_{xy}(t)e^{-F_{xy}(t)},
\end{align}
where $F_{xy}(t) = \ln W_{xy}(t)p_y(t) - \ln W_{yx}(t)p_x(t)$. In Property~V,
we proved that $I_F = -\bra{2}\dot{E}(t)\ket{p(t)}$. Using this relation,
together with the definition of the thermodynamic forces, shows the Fisher
information is given by
\begin{align}
  I_F(t) &= -\bra{2}\dot{E}(t)\ket{p(t)} \nonumber \\
  &=-\sum_{x,y}\left[\frac{d}{dt}\left(W_{xy}(t)e^{-F_{xy}(t)}\right)\right]p_y(t) \nonumber \\
  &= -\sum_{x,y}\Big( \dot{W}_{xy}(t)e^{-F_{xy}(t)}p_y(t) \notag\\
&\phantom{{}=\quad\quad\,} - W_{xy}(t)\dot{F}_{xy}(t)e^{-F_{xy}(t)}p_y(t)\Big) \nonumber \\
  &= -\sum_{x,y}W_{xy}(t)p_y(t) \frac{d}{dt}\ln\left(\frac{p_y(t)}{p_x(t)}\right) \nonumber \\
  &= \llangle \dot{I}(t)\rrangle.
  \label{eq:IFIdot}
\end{align} 
Since $-\ln p_y(t)$ is the surprisal of state $y$ at time $t$, the quantity
$\llangle \dot{I}(t)\rrangle$ is the time rate of change in the surprisal
difference (between $y$ and $x$).

To connect $I_F(t)$ to the entropic acceleration, we differentiate the
entropy rate,
\begin{equation}
  \dot{S}(t) = \sum_{x,y} W_{xy}(t)p_y(t) \ln\left[\frac{p_y(t)}{p_x(t)}\right],
\end{equation}
with respect to time:
\begin{align}
  \ddot{S}(t) &= -\sum_{x,y}\frac{d}{dt}\left[W_{xy}(t)p_y(t)\right]I_{xy} - \sum_{x,y}W_{xy}(t)p_y(t)\dot{I}_{xy}(t) \nonumber \\
  &= \sum_{x,y}\frac{d}{dt}\left[W_{xy}(t)p_y(t)\right]I_{xy}(t) - I_F(t) \nonumber \\
  &= -\sum_{x}\ddot{p}_x(t)\ln p_x(t) - I_F(t),
  \label{eq:ent}
\end{align}
where $I_{xy} = -\ln p_y(t) + \ln p_x(t)$. From Eq.~(\ref{eq:IFIdot}), the
second term in the entropic acceleration is minus the Fisher information.  The
entropy rate $\dot{S}(t)$ is an average over the change in information $I(t)$,
so $\ddot{S}(t)$ can be thought of as the rate of change of the bulk, or
average information, $\ddot{S}(t) = d\llangle I(t)\rrangle/dt$. The Fisher
information, though, is an average over the rate of change in information
between each set of states $x$ and $y$, $I_F = \llangle \dot{I}(t)\rrangle$.
Therefore, the first term on the right hand side of Eq.~(\ref{eq:ent}) is the sum
of the bulk information rate and the average local information rate,
\begin{align}
  -\sum_{x}\ddot{p}_x(t)\ln p_x(t) &= \ddot{S} + I_F \nonumber \\
  &= \frac{d\llangle I(t)\rrangle }{dt} + \bllangle \frac{d I(t)}{dt}\brrangle.
\end{align}

\section{Special cases}

An intuitive physical interpretation of the results in the main text is acquired by considering  several
special cases. The Fisher information is 
\begin{align}
  I_F \equiv \llangle \dot{I} \rrangle = \sum_{x,y}W_{xy}p_y \dot{I}_{xy}
  &= \llangle\ddot{S}^i\rrangle + \llangle\ddot{S}^e\rrangle.
  \label{eq:IForce}
\end{align}
It is important to note that the angled brackets on the outside of the time
derivative refer to the average over the local change, instead of the
change in the average quantity, i.e., $\ddot{S}^i \neq \llangle
\ddot{S}^i\rrangle$, in general. The average entropic accelerations $\llangle
\ddot{S}^i\rrangle$ and $\llangle \ddot{S}^e\rrangle$ are defined as
\begin{equation}
  \llangle \ddot{S}^i\rrangle = \sum_{x,y}W_{xy}(t)p_y(t)\frac{d}{dt}\ln \left(\frac{W_{xy}(t)p_y(t)}{W_{yx}(t)p_x(t)}\right)
\end{equation}
and
\begin{equation}
  \llangle \ddot{S}^e\rrangle = -\sum_{x,y}W_{xy}(t)p_y(t)\frac{d}{dt}\ln \left(\frac{W_{xy}(t)}{W_{yx}(t)}\right).
\end{equation}
The uncertainty relation can be written as
\begin{align}
  \int_{t_0}^{t_f} dt\,I_F
  = \int_{t_0}^{t_f} dt\, \left[\llangle\ddot{S}^i\rrangle + \llangle\ddot{S}^e\rrangle\right]
  \geq \frac{\sigma^{2}\tau}{2}.
  \label{eq:IFsOne}
\end{align}

\noindent \textbf{Steady-states.} In a steady state, $\dot{p}_x = 0$ and the
Fisher information is $I_F = 0$. As a consequence, there is zero path
uncertainty $\sigma^2= \mathcal{I} = 0$. From Eq.~(\ref{eq:IForce}), this means
that the local entropy production rate and entropy flow rate are exactly in
balance: $\llangle\ddot{S}^i\rrangle = \llangle \dot{F} \rrangle = -
\llangle\ddot{S}^e\rrangle$.\\

\noindent \textbf{Geodesic.} If the average local rate of information change is
constant, (i.e., the path is certain):
\begin{equation}
  \mathcal{I} = \llangle\dot{I}\rrangle =  \llangle\ddot{S}^i\rrangle + \llangle\ddot{S}^e \rrangle\quad\quad \text{(geodesic)}.
\end{equation}
The average local entropic acceleration is constant, independent of time and
distance from the final distribution. \\

\noindent \textbf{Zero entropy flow.} If the entropy flow is zero, say, when
the system is connected to an idealized bath~\cite{esposito2010three}, then the
uncertainty principle becomes
\begin{align}
  \tau^{-1}\int_{t_0}^{t_f} dt\, \llangle\ddot{S}^i\rrangle
  \geq \frac{\sigma^{2}}{2}.
  \label{eq:IFsTwo}
\end{align}

\noindent \textbf{Zero entropy production.} If the entropy production is zero,
then the uncertainty principle becomes
\begin{align}
  \tau^{-1}\int_{t_0}^{t_f} dt\, \llangle\ddot{S}^e\rrangle
  \geq \frac{\sigma^{2}}{2}.
  \label{eq:IFsThree}
\end{align}

\noindent \textbf{Exponential probability distributions.} While the above
results are independent of the form of the probability distributions, the
special case of an exponential energy distribution gives further physical
insight. Of particular interest is a single-component, homogeneous, closed
system at thermal equilibrium with a reservoir at inverse temperature $\beta =
1/k_BT$:
\begin{equation}
  p_x(t) = Z^{-1}e^{-\beta \epsilon_x(t)}.
\end{equation}
Using the detailed balance condition (Property II above), which holds
regardless of the distribution, the probability of state $y$ relative to state
$x$ is
\begin{equation}
  \frac{p_y(t)}{p_x(t)} = \frac{E_{yx}(t)}{E_{xy}(t)} = e^{-\beta \left[\epsilon_y(t)-\epsilon_x(t)\right]} = e^{-\beta q_{yx}(t)}.
\end{equation}
We define the energy exchanged as heat during the transition from $x$ to $y$ as
$q_{yx} = \epsilon_y(t)-\epsilon_x(t)$. With this definition, the Fisher
equation and the heat are directly related,
\begin{align}
  I_F(t) &= -\beta\llangle\dot{q}(t)\rrangle_{xy} = -\sum_{xy}W_{xy}(t)p_y(t)\dot{q}_{xy}(t) \nonumber \\
  &= -\beta\llangle\dot{\epsilon}(t)\rrangle = -\sum_{xy}W_{xy}(t)p_y(t)\dot{\epsilon}_{x}(t).
\end{align}
As a result, the Fisher information is the rate at
which energy flows into the system as heat, scaled by the temperature of the
reservoir. The time average of this quantity appears in the uncertainty
principle, the second main result in the paper. For a system evolving along the
geodesic, $-\beta\llangle\dot{q}(t)\rrangle_{xy}$ is independent of time: energy is
exchanged between the system and reservoir as heat at a constant rate.

For this special case, we have the Shannon information,
\begin{equation}
  S(t) = -\sum_x p_x(t)\ln p_x(t) = \beta\left<\epsilon(t)\right>,
\end{equation}
the entropy rate,
\begin{equation}
  \dot{S}(t) = \sum_{xy} W_{xy}(t)p_y(t)\ln \frac{p_y(t)}{p_x(t)} = \beta\llangle q(t)\rrangle_{xy},
\end{equation}
and the entropic acceleration,
\begin{align}
  \ddot{S}(t) &= \beta\frac{d}{dt}\llangle q(t)\rrangle_{xy} \nonumber \\
   &= \beta\sum_x \frac{d}{dt}\left[W_{xy}(t)p_y(t)\right]q_{xy}(t) + \beta\bllangle\frac{dq(t)}{dt}\brrangle_{\hspace{-1.2mm}xy} \nonumber \\
  &= \beta\sum_x \ddot{p}_x(t)\epsilon_x(t) + \beta\bllangle\frac{dq(t)}{dt}\brrangle_{\hspace{-1.2mm}xy}.
\end{align}
We now drop the subscript $xy$ on the average. Integrating from an initial
time, $t_0$, to a final time, $t_f$, shows that over any arbitrary interval of
time,
\begin{equation}
  \llangle q(t)\rrangle\big|_{t_0}^{t_f} = \int_{t_0}^{t_f}dt\sum_x \ddot{p}_x(t)\epsilon_x(t) + \int_{t_0}^{t_f}dt\llangle\dot{q}(t)\rrangle,
\end{equation}
there are two contributions to the energy exchanged as heat. We note that if
the heat flow is constant, as in a nonequilibrium steady-state, or zero, as in
an equilibrium state, then both sides of this equality must vanish. Also, we
can recognize the second term as the time-integrated (necessarily positive)
Fisher information up to a factor of $\beta$. In this case, the first
inequality in the main text is
\begin{equation}
  \frac{\beta}{\tau}\int_{t_0}^{t_f}dt\llangle\dot{q}(t)\rrangle \geq
  \frac{\sigma^2}{2}.
\end{equation}
Using this inequality, we get
\begin{equation}
  \llangle q(t)\rrangle\big|_{t_0}^{t_f} \leq \int_{t_0}^{t_f}dt\sum_x \ddot{p}_x(t)\epsilon_x(t) - \frac{\sigma^2\tau}{2\beta}.
\end{equation}

\noindent \textbf{Energy fluctuations.} Crooks, \cite{crooks2011fisher} showed
that for a canonical ensemble, the Fisher information is equal to the
infinitesimal change in energy with respect to a change in the control
parameter,
\begin{equation}
I_F(\theta) = \beta^2\left\langle\left(  \left\langle\frac{\partial \epsilon}{\partial\theta}\right\rangle - \frac{\partial \epsilon}{\partial \theta}\right)^2\right\rangle.
\end{equation}
If $\beta$ is the control parameter, this expression becomes the variance
measuring energy fluctuations around equilibrium. Here, we derive a similar
result. We will also assume that our distributions are exponential, but that
the control parameter (here, time) is varied smoothly and arbitrarily.
\begin{figure*}[!t]
\centering
\includegraphics[width=1\textwidth]{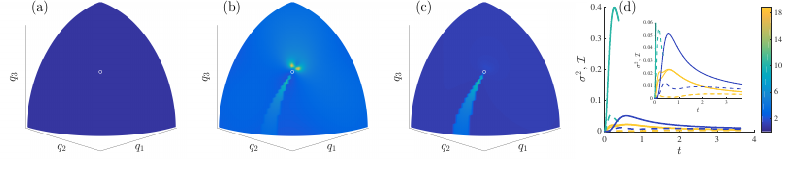}

\caption{\label{fig:sp} Scars on the state space (a)-(c) appear over time from
either higher information rates or low path uncertainties. (a) Initially, all
initial conditions have a low ratio of the cumulative rate of information
change, $\mathcal{I}$, and uncertainty, $\sigma^2$. (b) At intermediate times,
different initial conditions have a drastically different value of
$\mathcal{I}/\sigma^2$. (c) At long times, slow initial conditions are washed
out by time averaging but those scarring the state space reach the stationary
state (open circle) quickly. (d) The uncertainty and cumulative rate of
information change for four representative initial conditions. Green and blue
lines are initial conditions an equal distance from the stationary state
(colored according to the value of $\mathcal{I}/\sigma^2$). Yellow lines
correspond to the bright $\mathcal{I}/\sigma^2$ regions in (c).}
\vspace{-0.1in}

\end{figure*}
The distributions are then time dependent and $\beta$ is fixed. In this case,
$I_F$ measures the fluctuations in the energy rates. Given $p_x(t) =
Z(t)^{-1}\exp(-\beta\epsilon_x(t))$, the change in probability is,
\begin{align}
  \dot{p}_x(t) &= -\beta \dot{\epsilon}_x(t)p_x(t) - p_x(t)\frac{\dot{Z}}{Z}\nonumber \\
  &= -\beta \dot{\epsilon}_x(t)p_x(t) + \beta p_x(t)\sum_x \dot{\epsilon}_x(t)p_x(t).
\end{align}
Writing the ensemble average as $\sum_x p_x(t)[\cdot] = \langle \cdot\rangle$,
\begin{align}
  -\frac{d\ln p_x(t)}{dt} &= \beta \langle\dot\epsilon(t)\rangle - \beta \dot\epsilon_x(t), \nonumber \\
  \left(\frac{d\ln p_x(t)}{dt}\right)^2 &= \beta(\dot\epsilon_x(t) - \langle\dot{\epsilon}(t)\rangle)^2, \nonumber \\
  \sum_x p_x(t)\left(\frac{d\ln p_x(t)}{dt}\right)^2 &= \beta^2 \text{Var}[\dot{\epsilon}(t)], \nonumber \\
  I_F(t) &= \beta^2 \text{Var}[\dot{\epsilon}(t)].
\end{align}
In this case, the energy is playing the role of our control parameter and
$\beta$ is held fixed, so the Fisher information measures the fluctuations in
the energy rate. Using the first inequality from the main text gives,
\begin{equation}
  \frac{\beta^2}{\tau}\int_{t_0}^{t_f}dt\,\text{Var}[\dot{\epsilon}(t)] \geq \frac{\sigma^2}{2}.
\end{equation}
The uncertainty over the path lower bounds time-averaged fluctuations of the
energy rates. The geodesic is a path traversed when systems have no
fluctuations in the energy rate. All other paths will have a positive variance.

\section{Models}

\noindent \textbf{Landauer principle.} Consider a system initially exchanging
energy with the reservoir at inverse temperature $\beta$ and relaxing to
thermal equilibrium. The two states $x = \{1,0\}$ are observed with probability
$p_x(t_0) = [.5,.5]$ initially and $p_x(t_f) = [1,0]$ at the final time. The
net result is the erasure of state $x = 0$. So that the heat can be measured
during this evolution, we choose a driving protocol such that $p_x(t) =
Z(t)^{-1}e^{-\beta \epsilon_x(t)}$, where $Z(t)$ is the partition function,
$\beta$ is the inverse temperature of the bath, and $\epsilon_x(t)$ is the
energy at state $x$.

\begin{figure}[!b]
\centering
\includegraphics[width=.7\columnwidth]{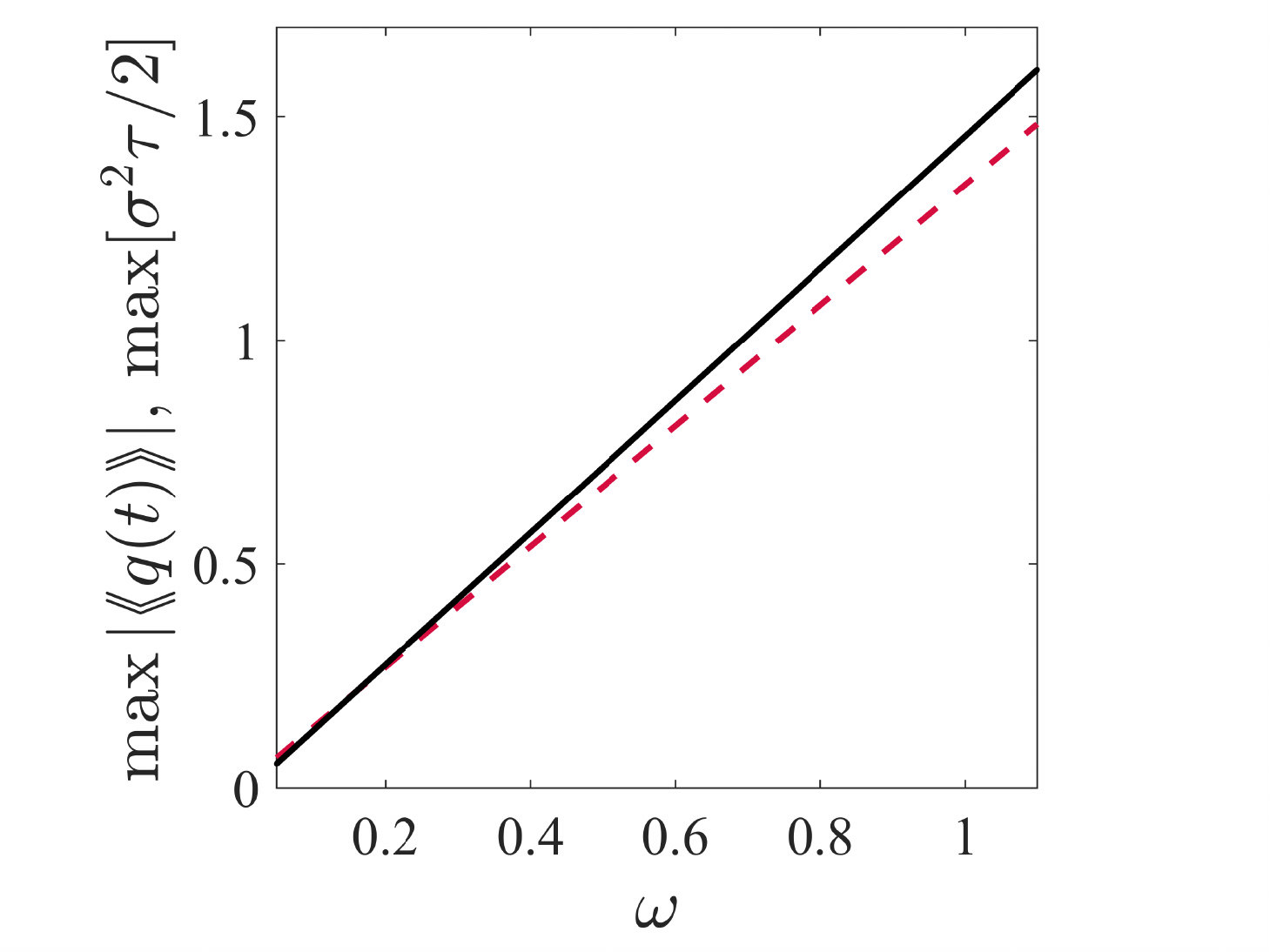}

\caption{\label{fig:maxHeat} The maximum instantaneous average heat and maximum
uncertainty grow linearly with increasing driving rate, $\omega$. The
coefficients for $\max|\llangle q(t)\rrangle| = a_1\omega + a_2$ are, $a_1 =
1.349$, $a_2 = -1.096\times 10^{-5}$, while for $\max[\sigma^2\tau/2]$ they
are $a_1= 1.477$ and $a_2 = -0.02$.}

\end{figure}

To ensure that our dynamics generates the prescribed sequence of distributions,
we work backwards. Given $p_x(t)$ for $t_0 \leq t \leq t_f$, we estimate
$\dot{p}_x(t) = [p_x(t+\delta t) - p_x(t)]/\delta t$, where $\delta t = 1\times
10^{-4}$ was used for all simulations. To find the $W(t)$ that satisfies
$W(t)\ket{p(t)} = \ket{\dot{p}(t)}$, we solve the system of linear equations
using conservation of probability, $W_{11}(t) = -W_{21}(t)$, $W_{22}(t) =
-W_{12}(t)$. There is an infinite family of $W(t)$ that satisfies these
conditions given by, 
\begin{equation}
W(t) =
  \begin{bmatrix}
  -A & \frac{\dot{p}_1(t) + Ap_1(t)}{p_2(t)} \\
  A &  -\frac{\dot{p}_1(t) + Ap_1(t)}{p_2(t)}
  \end{bmatrix}.
\end{equation}
We set $A$, which is arbitrary, to one. To calculate the second inequality in
the main text, we need to evaluate $-\sum_x \ddot{p}_x(t)\ln p_x(t)$. Given
$p_x(t) = Z(t)^{-1}e^{-\beta\epsilon_x(t)}$, the first and second derivatives
give,
\begin{equation}
  \dot{p}_x(t) = \beta p_x(t)\left(\dot{\epsilon}_2(t)p_2(t) - \dot{\epsilon}_x(t)\right).
\end{equation}
\begin{align}
  \ddot{p}_x(t) = &\beta\dot{p}_x(t)\left[\dot{\epsilon}_2(t)p_2(t) - \dot{\epsilon}_x(t)\right] + \nonumber \\
  &+ \beta p_x(t)\left[\dot{\epsilon}_2(t)\dot{p}_2(t) + \ddot{\epsilon}_2(t)p_2(t) - \ddot{\epsilon}_x(t)\right].
\end{align}
For the energies, we hold $\epsilon_1(t) = \epsilon_2(t_0)$ constant, making
our initial distribution $p_x(t_0) = [0.5,0.5]$, and vary $\epsilon_2(t) =
c_1 +\frac{c_2}{\pi}\atan(\omega t)$, where we use $c_1 = 0.2$, and $c_2 = 20$.
The first and second derivatives for $p_x(t)$ are then
\begin{equation}
  \dot{\epsilon}_2(t) = \frac{20\omega}{\pi[ 1+\omega^2t^2]},
\end{equation}
\begin{equation}
  \ddot{\epsilon}_2(t) = \frac{-40\omega^3 t}{\pi[1 + \omega^2t^2]^2}.
\end{equation}

In the main text, we describe how the system initially follows the geodesic and
only deviates after the maximum heat has been dissipated. We also find for this
system that both $\max|\llangle q(t)\rrangle|$ and $\max[\sigma^2\tau/2]$
increase linearly with $\omega$. For $20$ evenly spaced values of $\omega$,
ranging from $0.05$ to $1.1$ we evolve each for the same time, and record the
maximum value of $\llangle q(t)\rrangle$ and $\sigma^2\tau/2$. These quantities
are shown in Fig.~(\ref{fig:maxHeat}), with the maximum heat in red and maximum
uncertainty in black.\\

\begin{figure}[!ht]
\centering
\includegraphics[width=0.59\columnwidth]{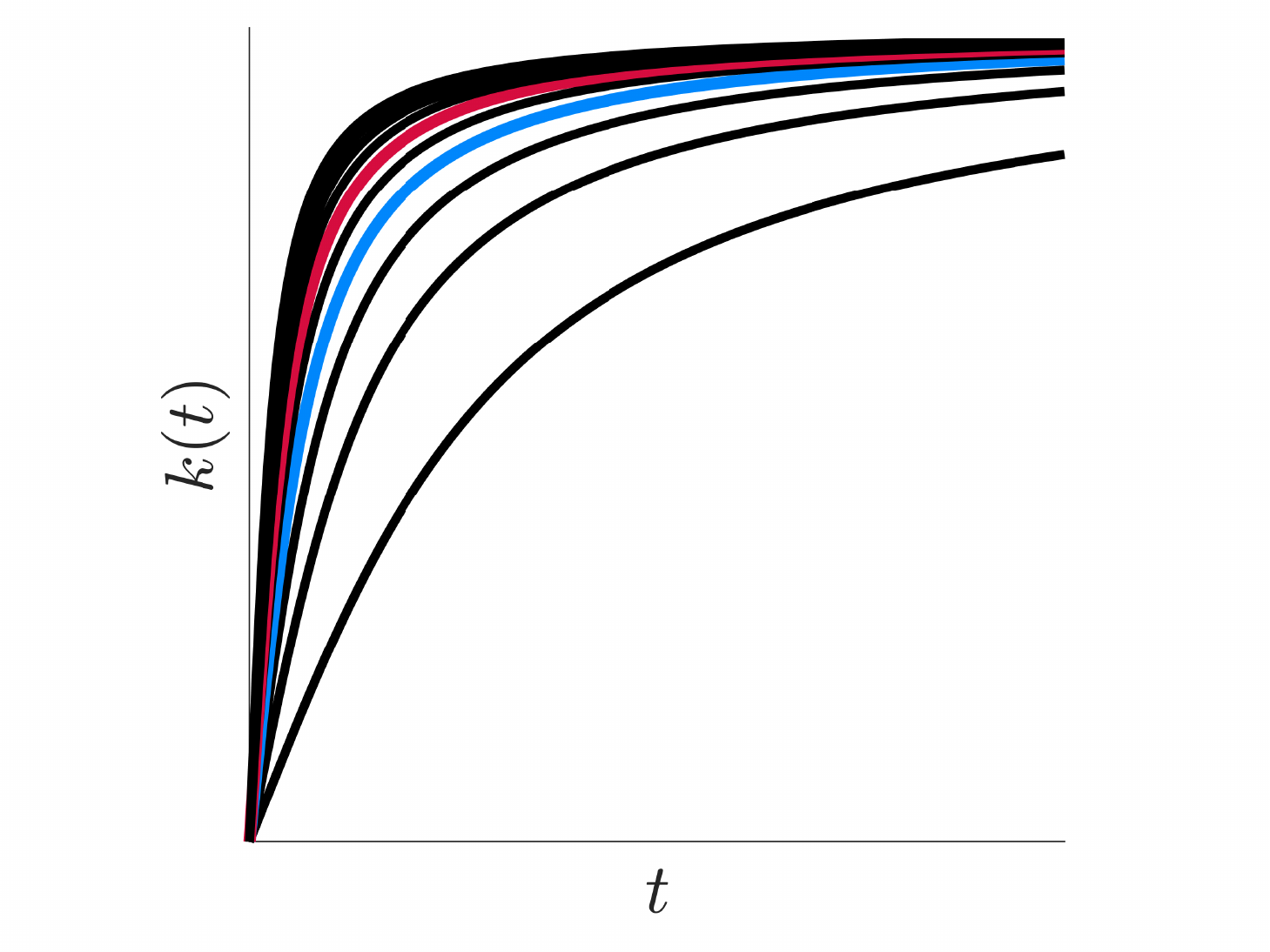}

\caption{\label{fig:RP} Rate coefficients for different values of the driving
rate $\omega$. The lines in blue and red correspond to $\omega = 4$ and $\omega
= 6$, respectively, used in the three state model.}

\end{figure}

\noindent \textbf{Driven versions of Onsager's three-state model} For the three
state model in the manuscript, the rate matrix takes the form: \[ W = 
\begin{bmatrix}
 -2k_t^- & k^+_t & k^+_t \\
 k^-_t   & -(k^-_t +k^+_t) & k^+_t \\
 k^{-}_t & k^{-}_t & -2k^{+}_t
\end{bmatrix}.
\]
The $k^+_t$ rate is given by $k^+_t = \frac{4}{\pi}\atan(\omega^+t)$ and the
$k^-_t$ rate is $ k^-_t = \frac{4}{\pi}\atan(\omega^-t)$ as shown in
Fig.~(\ref{fig:RP}).

\begin{figure}[t]
\centering
\includegraphics[width = 0.89\columnwidth]{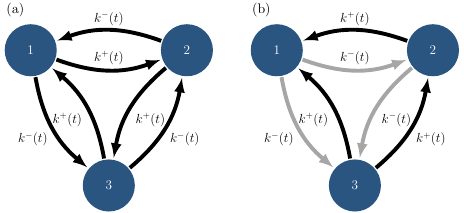}

\caption{Kinetic scheme for a driven version of Onsager's three-state model
with (a) symmetry and (b) broken symmetry.}

\end{figure}

As we see, increasing $\omega$ increases the rate that
$k$ approaches a limiting value. In this model, every initial condition reaches
the same stationary state $p^\infty = [1/3,1/3,1/3]$.  

To test the uncertainty relation, we calculate the ratio,
$\mathcal{I}/\sigma^2$ to confirm that it is larger than $1/2$,
Fig.~(\ref{fig:sp}a-c). Each initial condition starts with roughly the same
ratio of information to uncertainty, Fig.~(\ref{fig:sp}a). Because the rate of
information change and the variance are path dependent, the dynamics quickly
generate a wide range of ratios, Fig.~(\ref{fig:sp}b). One prominent feature is
the two (yellow,~\footnote{Though starting from different initial conditions,
$q(t_0) = [0.5539,0.5470,0.6277]$, $q(t_0) = [0.5213,0.5989,0.6079]$, both
regions have equal uncertainty at $t = 0.9096$ and (numerically)
equivalent values of $\sigma^2$ and $\mathcal{I}$ at all times}) regions of
initial conditions that have a large ratio relative to the rest of the state
space. The time series of these initial conditions are shown in
Fig.~(\ref{fig:sp}d) (yellow), where we see that it is predominately the small
uncertainty that is generating the large ratio. We also see that this large
ratio disappears as the path approaches $p^\infty$.

Fig.~(\ref{fig:sp}c) shows the ratio for the (minimum) time it takes for all
initial conditions to become within $5\times 10^{-3}$ of $p^{\infty}$. Also
prominent is a ``scar'' where initial conditions quickly close in on
$p^{\infty}$ at the cost of high cumulative entropic accelerations/change in
information. Fig.~(\ref{fig:sp}d) shows the time series for one of these
initial conditions (green solid/dashed lines $q(t_0) =
[0.6491,0.5796,0.4927]$). To contrast paths originating from the scar, the blue
line in Fig.~(\ref{fig:sp}d) corresponds to an initial condition ($q(t_0) =
[0.4953,0.5751,0.6511]$) that is equidistant from $p^\infty$ but lies outside
the scar. The path followed from this starting point has a lower $\sigma^2$ and
$\mathcal{I}$, showing that the shortest distance from the stationary point is
a poor indicator of the behavior of the process.

\begin{figure*}[!ht]
\centering
\includegraphics[width=.9\textwidth]{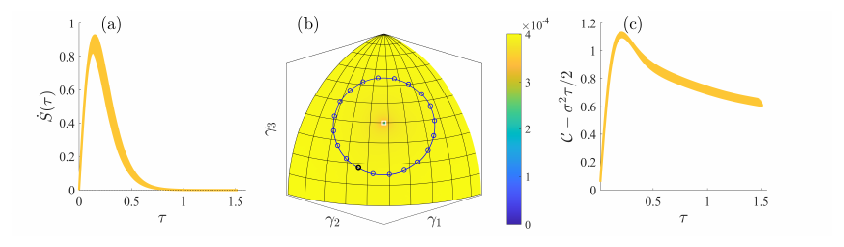}

\caption{\label{fig:ESym} Lack of state-space scarring in the symmetric, driven
Onsager model. (b) Each point on the octant of the sphere is an initial
condition that we evolve to the stationary distribution $p^\infty =
[1/3,1/3,1/3]$. Unlike the asymmetric case, each initial condition generates
the same final value of $\dot{S}(\tau)$. (a) The entropy rate $\dot{S}(\tau)$
for the set of conditions specifed by the blue circles. (c) The upper bound of
these points, all of which evolve similarly in time.}

\end{figure*}

In the main text, we analyze a driven three state model with broken
symmetry. For comparison, we show the symmetric driven Onsager model. The
same driving protocol is used but the rate matrix is
\[
W = 
\begin{bmatrix}
  -(k^-_t +k^+_t)  & k^-_t & k^+_t \\
 k^+_t   & -(k^-_t +k^+_t) & k^-_t \\
 k^{-}_t & k^{+}_t &  -(k^-_t +k^+_t)
\end{bmatrix}.
\]
There is a forward cycle and a reverse cycle. We see in Fig.~(\ref{fig:ESym})
that there is no scarring from high entropy rate initial conditions. Now every
initial condition reaches approximately the same final entropy rate in the same
time $\tau$. Initial conditions an equal distance from the stationary point
seem to have nearly equivalent time evolutions, varying only slightly in the
final, maximum value of $\dot{S}(\tau)$ reached. The upper bound (on
$\dot{S}(\tau)$) also has nearly the same temporal behavior, regardless of
initial condition. It appears that the breaking of the symmetry in the rate
matrix leads to the scarring of state space and possibility of dissipative
initial conditions.

\end{document}